\documentclass[12pt]{article} 
\input{epsf.tex}
\begin{document}
\begin{center}
{\Large \bf 
Cascade population of levels and probable phase transition in vicinity of the 
excitation energy $\approx 0.5B_n$ of heavy nucleus}
\end{center}

\begin{center} V.A. Bondarenko$^1$,
J. Honz\'atko$^2$, V.A. Khitrov$^3$, Li Chol$^3$,  Yu.E. Loginov$^4$,\\
 S.Eh. Malyutenkova$^4$,
A.M. Sukhovoj$^3$,
I. Tomandl$^2$,
\end{center}
\begin{center} {\it $^1$Institute of Solid State Physics, University of Latvia, LV 2169\\
 $^2$Nuclear Physics Institute, CZ-25068, \v{R}e\v{z} near Prague, Czech
Republic\\
  $^3$FLNP, Joint Institute for Nuclear
Research, Dubna, Russia\\
$^4$Peterburg Nuclear Physics Institute, Gatchina, Russia}
\end{center}

From the comparison of absolute intensities of the two-step $\gamma$-cascades  and 
known intensities of their primary and secondary transitions, the cascade and 
total population of about $\sim 100 $ levels of $^{181}$Hf and $^{184,185,187}$W
excited in thermal neutron capture was determined.
These experimental results and intensities of two-step cascades
to the low-lying levels of mentioned  nuclei were reproduced in calculation 
using level densities with clearly expressed step-like structure. Radiative strength 
functions of the primary transitions following $\gamma$-decay of these compound 
nuclei to  the levels in the region of pointed structure are considerably enhanced. 
Moreover, population of levels below 3 MeV can be reproduced only with accounting 
for local and rather considerable increase in  radiative strength functions
of the secondary transitions to the levels in vicinities of break points in energy
dependence of level density and significant decrease of that to lower-lying 
states.

Simultaneous change in both level density and strength functions in the same
excitation region of a nucleus corresponds to the definition of the second-order 
phase transition.

\section{Introduction}\label{sec:intro}\noindent

Sufficient progress in understanding of the processing occurring in a nucleus at
 the excitation below the neutron binding energy was achieved from analysis [1] of intensities
           \begin{equation}
 I_{\gamma\gamma}=F(E_1)=\sum_{\lambda ,f}\sum_{i}\frac{\Gamma_{\lambda i}}
 {\Gamma_{\lambda}}\frac{\Gamma_{if}}{\Gamma_i}=\sum_{\lambda ,f}
 \frac{\Gamma_{\lambda i}}{<\Gamma_{\lambda i}>
 m_{\lambda i}} n_{\lambda i}\frac{\Gamma_{if}}{<\Gamma_{if}> m_{if}}
\end{equation}
of the two-step $\gamma$-cascades in function of the primary transition energy
obtained within the method [2]. This analysis was carried out for a group of nuclei 
from $^{28}$Al to $^{200}$Hg for which the coinciding $\gamma$-quanta following thermal neutron 
capture were measured. Comparison of the experimental and calculated intensities
of the cascades with the total energy $E_1+E_2=B_n-E_f$ (with energy of final levels 
$E_f<1$ MeV) for 51 nuclei shown that existing notions of the cascade $\gamma$-decay
process need very serious correction. There is no other way to improve accuracy 
of model calculation of the cascade $\gamma$-decay parameters of a nucleus.

Qualitative interpretation of the all totality of the data obtained in investigations
 of two-step cascade testifies that structure of the wave functions of the 
levels noticeably differs for the excitation regions above and below $0.5B_n$.
In the frameworks of modern theoretical notions of energy dependence of nuclear 
level density, this change is related with the process of breaking of paired nucleons. 
As a consequence, both level density and probability of their excitation (de-excitation)
 seriously differs from the predictions of models considering nucleus
 as a pure fermion system (for instance, [3,4]).

Significance of this conclusion follows from the fact that so simple models are 
used up to now for both analysis of the experiment and calculation of $\gamma$-ray 
spectra and neutron cross-sections of interaction with nuclei. Specific of the 
data obtained in [1], however, requires not only transition to more realistic 
models of level density $\rho$ and radiative strength functions
 $k=\Gamma_{\lambda i}/(E_{\gamma}^3\times A^{2/3}\times D_{\lambda})$
(like [5,6] and [7], respectively) but also more precise their parameterization
and further development. The necessity of further theoretical development follows not only 
from unsatisfactory correspondence between the model notions and experiment but 
also from a need in more precise interpretation of processes occurring in  nucleus.
Development of new methods of analysis of existing experimental data is 
oriented to this goal, as well.
\section{
New possibilities of the experiment}\noindent

A bulk of information on the intensities 
 \begin{equation}
i_{\gamma\gamma}=i_1 \times i_2/\sum i_2,
\end{equation}
experimentally resolved individual cascades was obtained for all 51 studied nuclei.
 Their parameters, including most probable quanta ordering, were derived from
 the experiment up to the excitation energy of the cascade intermediate levels 3
-5 MeV by means of the original method of analysis created in Dubna. One of the 
important element of this method is the quantitative algorithm [8] of 
improvement of energy resolution without decrease in efficiency in the spectra with equal 
total energy $E_1+E_2=const$.

Experiments in Riga and  \v{R}e\v{z} allowed measurements [9-11] of most complete spectra
of intensities of the primary $i_1$ and secondary $i_2$ $\gamma$-transitions up to the 
neutron binding energy $B_n$ in $^{181}$Hf and even-odd isotopes of W following thermal
neutron capture in corresponding target nuclei. Analogous data for $^{184}$W were 
obtained in Gatchina [12]. The data on $i_1$ and $i_2$ are available and for some 
other nuclei but their poor statistics do not allow one to repeat in full 
measure analysis described below (except $^{118}$Sn).
\section{
Method of analysis}\label{sec:meth}\noindent

From the eq.~(2) for the totality of the data on $i_{\gamma\gamma}$, $i_1$  and $i_2$
one can determine
the total population $P=\sum i_2$ for about 100 levels up to the excitation energy
of 3-4 MeV and higher. Difference of $P$ and intensity of primary transition $i_1$ to 
each of these levels is equal to sum of their population by two-step, three-step
and so on cascades. It can be calculated within different assumptions  and 
models of level density excited at the thermal neutron capture and radiative
strength functions of cascade $\gamma$-transitions (including $\rho$ and $k$ values obtained
according to [1]).

The region of the maximum discrepancy between the experiment and different 
variants of calculation shows where and in what direction should be modified model
description of the cascade $\gamma$-decay process.

It is clear that at present it is impossible to determine population of all
without exclusion intermediate levels of two-step cascade even at low excitation 
energy (owing to the detection threshold of intensities $i_{\gamma\gamma}$, $i_1$
and $i_2$). Therefore,
it is worth while to compare experiment and calculation for values of $P-i_1$ 
summed over small excitation energy intervals and consider these sums as the 
lower estimates for each of intervals. Comparison of this kind was carried out 
by us for different nuclei. Results of this analysis and other experimental data for
compound nuclei $^{181}$Hf and $^{184,185,187}$W are presented below (for them there was 
stored maximum information of needed type).

A degree of discrepancy between the calculated population $P-i_1$ and its unknown 
experimental value is determined by both incompleteness of the data on intensities
 of cascades and single transitions and possible strong influence of the structure 
 of the wave function of the excited level on probability of its cascade 
 population (such influence for the levels with the excitation energy higher than 2-3
 MeV is confirmed by strong fluctuation of neighbor levels). 

The primary dipole transitions of cascades excite in the considered even-odd 
isotopes the levels with $J^{\pi}=1/2^{\pm},3/2^{\pm}$,
and in $^{184}$W - levels of both parities in 
the spin window 0-2. Cascade population of these levels is determined not only by the 
averaged intensity of the ended by them cascades but also intensity of three-step,
 four-step an so on cascades. The latter appear at depopulation of levels from
 wider spin window. Therefore, it seams insufficient to relate large dispersion 
of population of the cascade intermediate levels only with difference in their spins.
\section{
Systematic errors of population of levels}\noindent

The minimum possible systematic error of level population can be achieved only 
at the maximum possible positive correlation of systematic errors which, 
according to eq.~(2), determine its magnitude. I.e., when absolute intensities $i_{\gamma\gamma}$ are
determined [14] using their relative values and absolute intensities $i_1\times i_2/\sum i_2$ 
calculated using the data like those listed in [9-12] for several cascades 
proceeding through the lowest-lying levels.  Decay scheme of these levels is usually 
well established and, correspondingly, the $\sum i_2$ value has minimum error. 
In this approach of normalization of the experimental data, the magnitude and sign of 
the error $\delta i_{\gamma\gamma}$ of intensity of any cascade strongly correlates with the 
mean error of intensity $\delta i_1$ of its primary transition, and the total relative 
error of population  $P-i_1$ is mainly determined by the total relative error of the $i_2$ values.

In modern experiment [9-12,13] the error of values $i_1$ and $i_2$ is mainly determined
by the error [15] of capture cross-section of thermal neutrons in the isotope
under investigation. In majority of cases it does not exceed 5-10\%.  Therefore,
the level population from eq.~(2) has a precision determined only by systematic
errors in the data [9-13] and random errors of concrete values $i_1$ and $i_2$. The 
only problem in determination of $P-i_1$ is unresolved doublets of the cascade 
secondary transitions $i_2$ and (to less extent due to less density of peaks in the 
spectra) primary transitions $i_1$. Partially the multiplets can be identified and 
resolved by approximation of single spectra of HPGe detector using information on
the two-step cascades and known decay scheme of the nucleus under consideration. 
These data should be obtained in the experiment on the thermal neutron capture 
in the highly enriched target. In other cases intensity $i_2$ can be distributed 
between the cascades in which such multiplet is proportional to $i_{\gamma\gamma}$ or they can be 
excluded procedure of determination of $P$ in plenty of the data on $i_{\gamma\gamma}$, $i_1$ and $i_2$ 
for the same intermediate level of cascade.

Some data for the nuclei considered here are listed in Table.

Table. {\it ~$N_{2\gamma}$ is the probable total number of levels populated by the primary E1 
and M1 transitions below 3 MeV in even-odd nuclei and 4 MeV in $^{184}$W estimated 
for the level density shown in Fig.~4; $N_c$ is the experimental number of the
resolved two-step cascades, $N_i$ is the number of levels for which the cascade
population was determined and portion (percent) of multiplets of secondary
transitions with multiplicity M. $\sum i_1$ and $\sum i_2$ are the total intensities (\% per decay)
of primary and secondary transitions of cascades, respectively, involved in analysis.
 $E_f$ is the energy of the cascade final level, and $d=\sum i_\gamma \times E_\gamma/B_n$ is the part of the 
total $\gamma$-spectrum observed in [9-12].}

\begin{center}
\begin{tabular}{|l| r |r| r| r| r |r |r| r| r|}
\hline
Nucleus     & $N_{2\gamma}$ &$N_c$ & $N_i$ & $(M=2)$& $(M\geq 3)$& $\sum i_1$, \% & $\sum i_2$, \%
& $E_f$, keV & $d$\\
\hline
$^{181}$Hf& 260&   69(58)   & 61    &  $\leq 14$  & 0          & 31.6 & 35.3 &  332 & 0.63\\
$^{184}$W & 135& 105(104)   & 78    &  $\leq 16$  & $\leq 2 $  & 55.3 & 107.2&  364 & 0.82\\
$^{185}$W & 156& 150(136)   & 135   &  $\leq 24$  & $\leq 25$  & 53.3 & 74.8 &  1068 & 0.60\\
$^{187}$W & 200& 121(120)   & 112   &  $\leq 25$  & $\leq 17$  & 41.2 & 65.6 &  303 & 0.47\\
\hline
\end{tabular}
\end{center}
{\it Exceeding of $N_c$ over $N_i$ is caused by  both better background conditions of 
registration of cascades to the low-lying levels as compared with the experiment 
on determination of $i_1$ and $i_2$ and the use of the of improving energy resolution [8].
 Table does not include cascades with excitation energy of intermediate levels 
higher than 3-4 MeV for which $i_1$ and $i_2$ are unknown. In the brackets for $N_c$ 
are given values of this parameter corresponding to $E_{ex} \leq 3$ MeV 
(or 4 MeV for $^{184}$W).}
\section{
Results}\label{sec:err}\noindent

Comparison between the obtained population of the excited states were performed 
in two variants:

(a)~the population of each from $N_i$ intermediate levels (including intensity of 
the primary transition populating it) was compared (Fig.~1) with 
several variants of the calculation;

(b)~the total cascade population $P-i_1$ corresponding to the 200 keV interval of
the excitation energy was compared (Fig.~2) with the same calculation.

 A necessity in two these variants follows, first of all, from the presence of
the detection threshold of intensities that limits a possibility to get information
on all levels of the studied nucleus. In the second, owing to the presence of
 systematic errors in determination of $i_1$ and $i_2$, the $P-i_1$ value is negative in 
some cases (this is possible due to low cascade population of level as compared 
with $i_1$). But the calculated total population $P$ of levels depends on the model
predicted values of level density and radiative strength functions weaker than
cascade population $P-i_1$. This is caused by the inevitable compensation of an
influence on population of change in, for example, $\rho$ by corresponding change 
in $k$ to the cascade primary transitions. That is why, it is necessary to get
additional confirmation for considerable discrepancy between the experimental
and calculated sums of $P-i_1$. This is provided by analogous comparison of the
experimental and calculated total populations of individual levels. 

\newpage
\begin{figure}
\vspace{4cm}
\leavevmode
\epsfxsize=16.5cm

\epsfbox{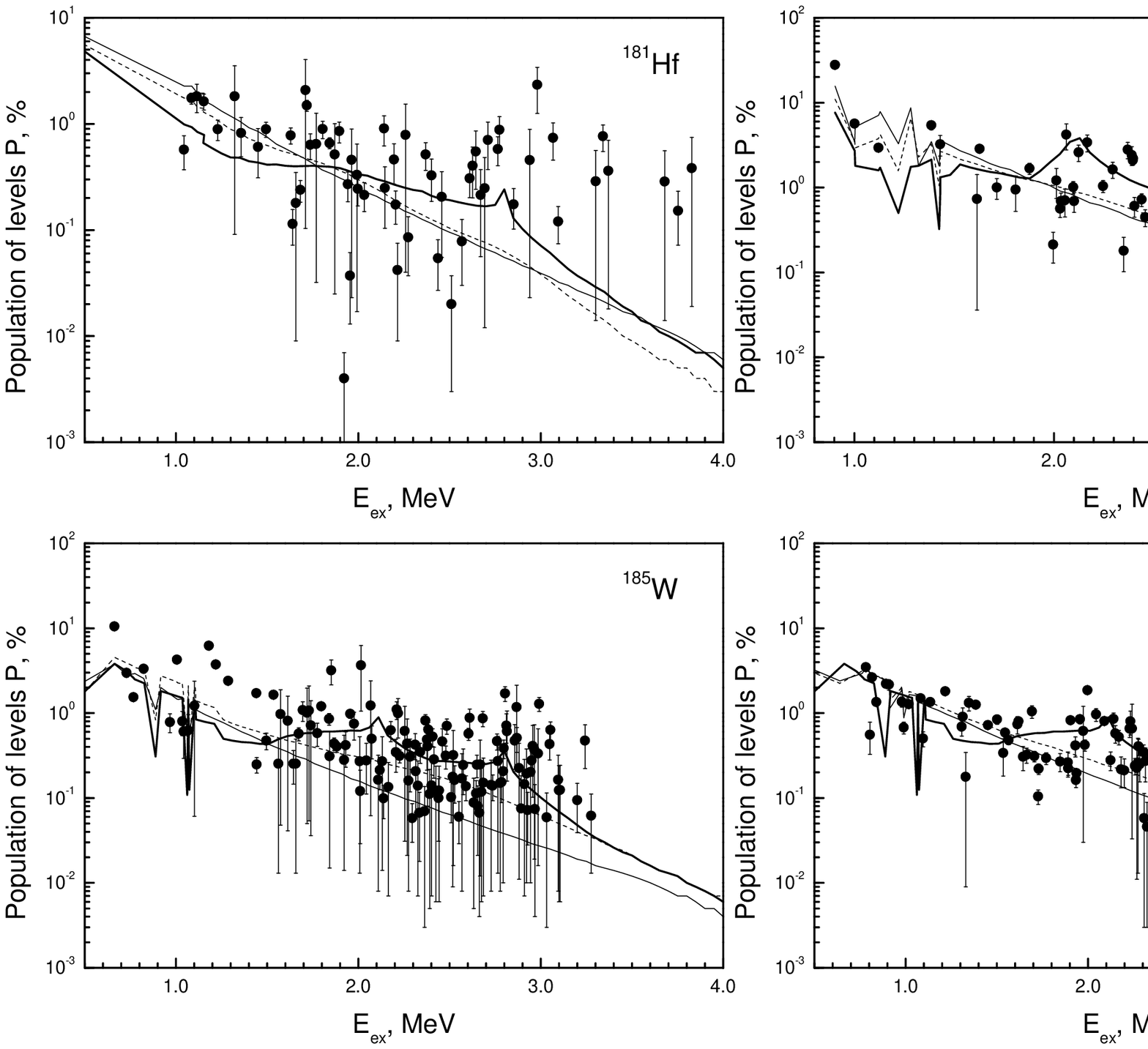}
\vspace{-6cm}
\end{figure}

{\bf {Fig. 1.} \it ~The total population of intermediate levels of two-step cascades (points
 with bars), thin
line represents calculation within models [3,4].
Dashed  line shows results of calculation using data [1]. Thick line shows results of 
calculation using level density [1], and corresponding strength functions of secondary
transitions are multiplied by function $h$ set by equations (3) and (4).}\\

The number of available for calculation variants of energy dependence of strength
functions and level density is large enough. But general rules of of change in
population of levels with the change in their excitation energy can be arrived 
with the use of only three variants of calculation:

(a)~level density is taken according to any model of non-interacting Fermi-gas; 
strength function of E1 transitions is set by known extrapolations of the Giant 
electric dipole resonance in the region below $B_n$, $k(M1)=const$ with normalization
 of $k(M1)/k(E1)$ to the experiment;

(b)~one can use $\rho$ and $k$ providing precise reproduction [1] dependence of the
two-step cascade intensities on the primary transition energy;

(c)~one can select a set of $\rho$ and $k$ which simultaneously reproduce $I_{\gamma\gamma}=F(E_1)$, 
$\Gamma_\gamma$ and maximum precisely reproduce $P-i_1$.
\newpage
\begin{figure}
\vspace{4cm}
\leavevmode
\epsfxsize=16.5cm

\epsfbox{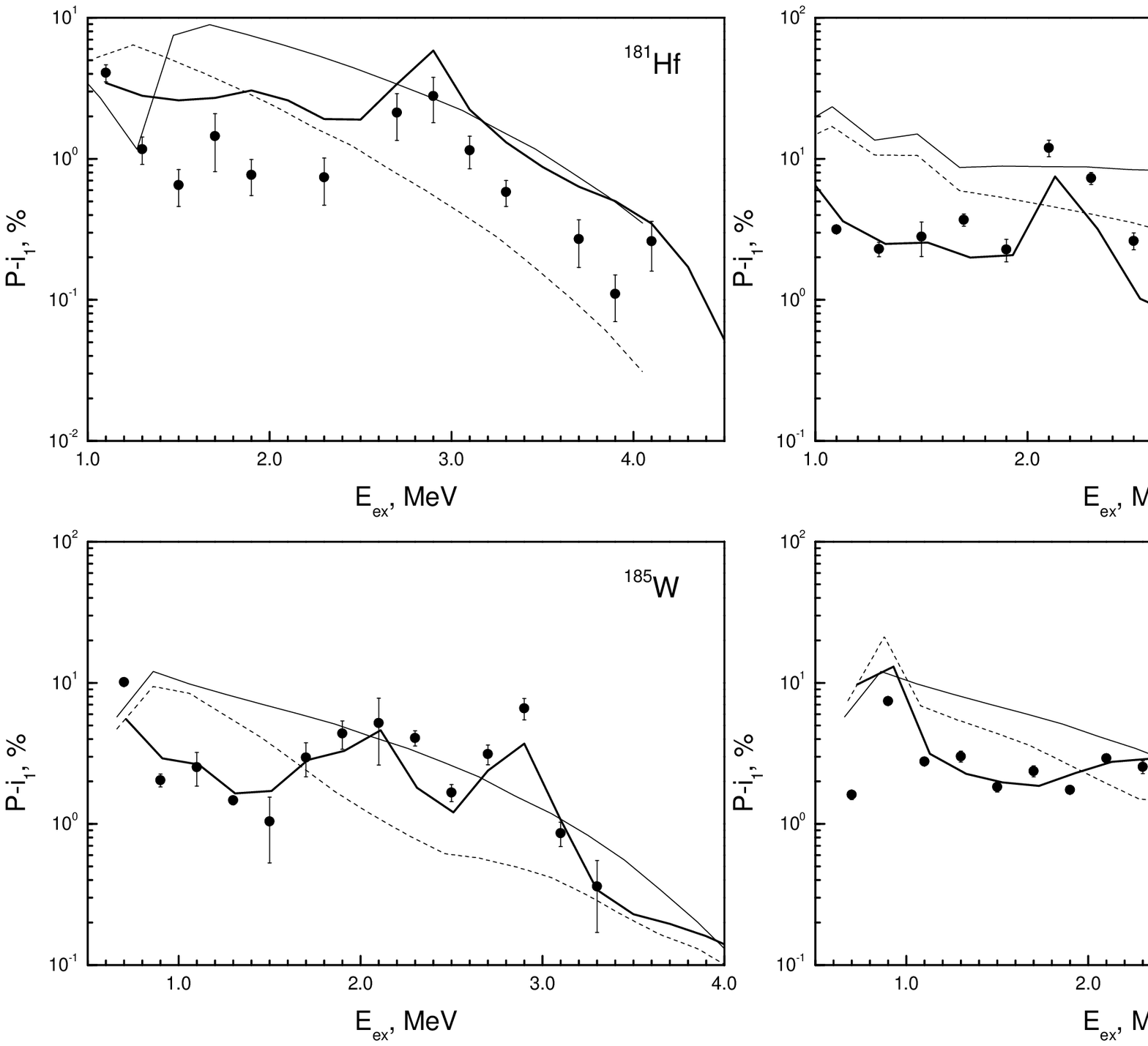}
\vspace{-6cm}
\end{figure}

{\bf {Fig. 2.}\it ~The same as in Fig.~1 for the total cascade population of levels in the
 200 keV energy bins.}\\

Variant (c) can be realized in iterative process: strength function $k$ of the 
secondary transitions obtained according to [1] are changed in the way providing 
better reproduction $P-i_1$. It is enough for this to multiply strength functions of 
secondary transitions to the levels lying below some boundary energy by the
function $h$ which contains few narrow enough peaks. Energy dependence of the form of 
these peaks can  be determined by analogy with the specific heat of macro-system
 in the vicinity of the second order phase transition as:
\begin{equation}
h=1+\alpha\times ({ln(|U_c-U_1|)-ln(|U_c-U|)})~~~in~the~case~of~~U<U_c,
\end{equation}
\begin{equation}
h=1+\alpha\times ({ln(|U_c-U_2|)-ln(|U_c-U|)})~~~in~the~case~of~~U>U_c,
\end{equation}
with some parameters $U_1$, $U_2$, and $U_c$.
The condition $(U_c-U_1) \neq (U_2-U_c)$ provides
required asymmetry of peaks and some more precise reproduction of cascade population
of levels at the "tails" of peaks as compared with the Lorentz curve, for instance.

In the best variant tested by us, their amplitude $\alpha$ increases
from zero at $U=B_n$ to maximum possible value (shown in figs.~4, 5)
when the excitation energy $U$
decreases. Positions of the peaks, their amplitude and shape are quite
unambiguously determined by values $P-i_1$. The correction functions found in this way are 
then included in analysis [1] to determine $\rho$ and $k$ providing correct reproduction
of cascade intensity.
\newpage
\begin{figure}
\vspace{4cm}
\leavevmode
\epsfxsize=16.5cm

\epsfbox{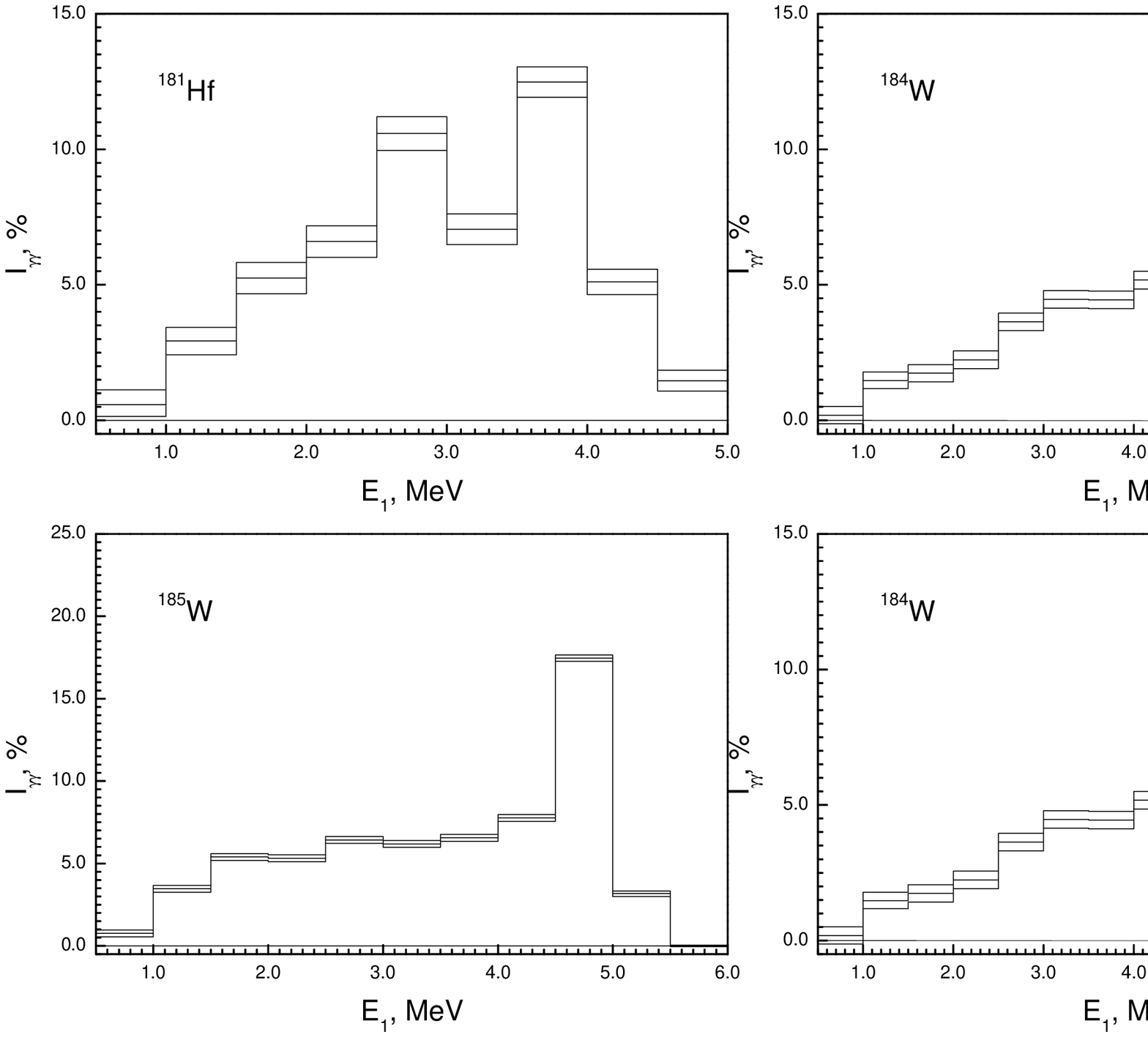}
\vspace{-6cm}
\end{figure}

{\bf {Fig. 3.}\it ~The intensity of the two-step cascade in function of the energy of their
 primary transitions.}\\\\
 The cascade intensities are shown in Fig.~3 and 
corresponding level densities and radiative strength functions together with the
functions $h$ are presented in figs.~4 and 5. Then this cycle is repeated one time when
we use the hypotheses of linearly increasing distortions in values $k(E1)$ and 
$k(M1)$ as decreasing the energy of decaying levels and several times  --- for the 
hypotheses $\alpha=const$. For minimization of the fitted parameters it was assumed 
that the correcting functions (figs.~4, 5) are equal for both electric and 
magnetic $\gamma$-transitions.

For all of the nuclei under consideration, the maximum value of $h$ is observed in
 vicinity of transition of level density from the practically constant value 
(the region of step-like structure) to the region of practically exponential increase.
Besides, there is observed correlation of maxima $k(E1)+k(M1)$ and $h$ with further
simultaneous decrease in their values when $E_1$ increases. This effect can be 
explained only by considerable influence of structure of level on strength
functions of populating it $\gamma$-transition. Moreover, one can assume such influence 
over whole interval of the levels excited in the $(n,\gamma)$ reaction.

Large enough number of hypothesis used for solution of this problem is inevitable.
But and in this case all conclusions about the $\gamma$-decay process of compound
nuclei should be considered rather as qualitative than as quantitative. So,
the presence of the clearly expressed ``step-like structure" in level density and
corresponding increase of $k(E1)+k(M1)$ (Fig.~5)can be considered with high probability
as established.
\newpage 
\begin{figure}
\vspace{4cm}
\leavevmode
\epsfxsize=16.5cm

\epsfbox{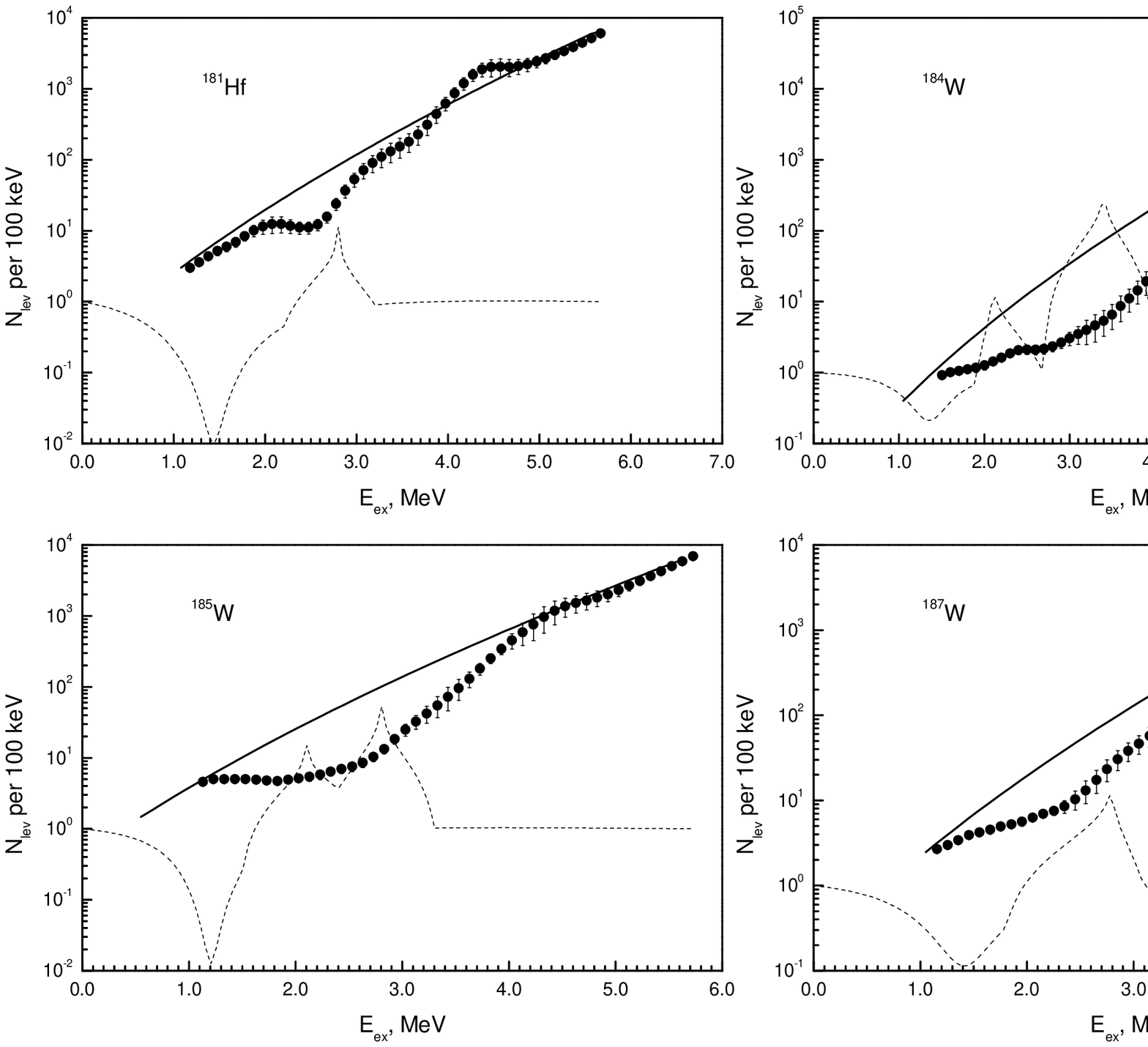}
\vspace{-6cm}
\end{figure}

{\bf{Fig. 4.} \it  ~The number of intermediate levels of two-step cascades in the case of 
different functional dependence of strength functions for  primary and secondary 
cascade transitions. Dashed line shows values of function $h$ for excitation energy
 $B_n-E_1$. Solid line represents predictions according to model [4].}\\\\\\
But the number and shape of these "steps" can be revealed 
most probably in future experiments. The same should be said about parameters of 
correcting functions $h$. If in region of excitations corresponding to considerable
increase of $k$ for the second, third and so on cascade is no doubts then concrete
parameters of function $h$ most probably should be considered as preliminary. 
They should be used for planning further experiments. In this respect is the
region $h<1$. It is impossible to conclude whether the strength of $\gamma$-transitions
is redistributed from the lower-lying levels excited by them to the higher-lying 
or this structure of $h$ caused only increase of $k(E1)+k(M1)$ in Fig.~5. But it
should be noted that observed values of $P-i_1$ for the considered nuclei could not be
reproduced without noticeable decrease of $k$ for the secondary transitions at low excitations.

Even the circumstance that we obtained only the lower estimation for $P-i_1$ cannot
be possible explanation for this problem. At the decrease of the excitation energy,
the portion of the unobserved population must decrease. It should be noted 
that the possibility of existence of specific dependence of product $k \times h$ does not
contradict the studied theoretically [6] regularity of fragmentation of simple 
state with any structure over nuclear levels. One of the most important qualitative
conclusion of this analysis is that the strength of this state can  concentrate
in asymmetric peaks with "tails" in the region of high excitation energy.
\newpage
\begin{figure}
\vspace{4cm}
\leavevmode
\epsfxsize=16.5cm

\epsfbox{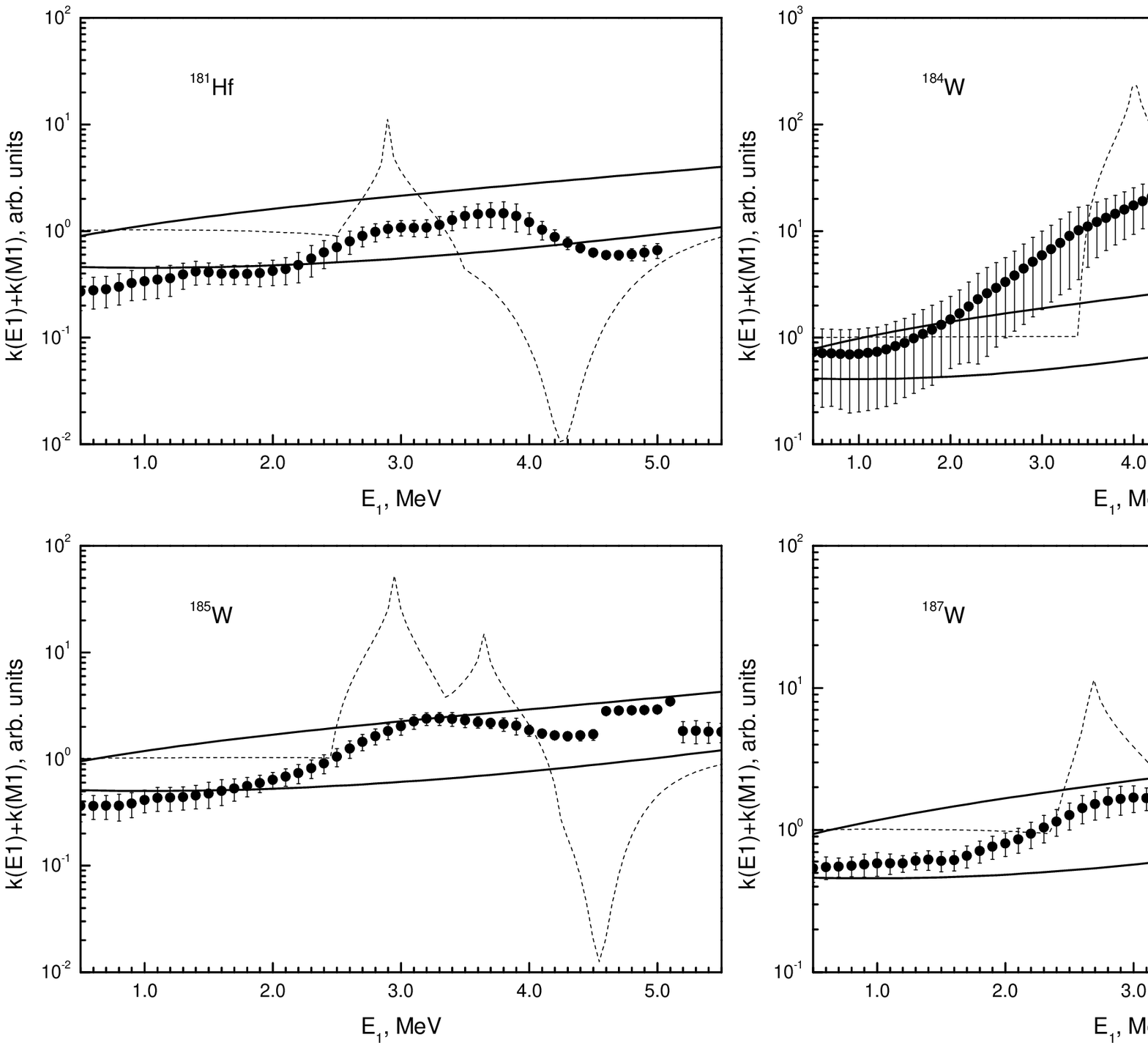}
\vspace{-6cm}
\end{figure}

{\bf{Fig. 5.} \it ~The sums of radiative strength functions of the cascade primary dipole
transitions providing reproduction of cascade intensities with the considered 
difference of their values with strength functions of secondary transitions 
(multiplied by $10^9$). Dashed line shows values of function $h$ for excitation energy $B_n-E_1$.}\\

The populations determined according to [1], density of the excited levels and 
parameters of strength functions depend the fluctuations of intensities of the 
primary transitions to the ground and low-lying excited states ($U<0.5$ MeV) which are 
not taken into account in this method. These transitions were not observed in 
cascades owing corresponding detection threshold and, therefore, only their mean values 
were included in calculation. This can increase discrepancy between $\rho$ and $k$ 
shown in figs.~4, 5 and corresponding model predictions. This is of maximum importance
for $^{181}$Hf where intensity of direct transition to the ground state is 23\% [13].
\section{Possible changes in model notions about the cascade $\gamma$-decay
process of compound nuclei}\noindent

According to method [1], the region of $\rho$ and $k$ values allowing reproduction of
 the experimentally obtained function $I_{\gamma\gamma}(E_1)$ with parameter $\chi^2/f<<1$ has been
 determined. Corresponding results for $^{181}$Hf and $^{185,187}$W  were published in [1] 
and [17], respectively. Estimation of influence of most important source of
systematic error on parameters of the cascade $\gamma$-decay were published in [18]. As
for the nuclei studied earlier, level density required for description of 
$I_{\gamma\gamma}(E_1)$ is noticeably less than predictions of the model of non-interacting Fermi-gas.
This level density qualitatively corresponds to developed by A.V.Ignatyuk and 
Yu.V.Sokolov notions of step-like energy dependence of level density what is the
result of breaking of Cooper pair (or several pairs [5]) of nucleons at
corresponding excitation energy of a nucleus.

In the frameworks of the generalized model of superfluid nucleus [5], level 
density above the phase transition from the superfluid to normal state is mainly 
determined by many-quasiparticle excitations. Below this energy, the properties
of a nucleus are strongly affected by boson branch of nuclear excitations. One can 
accept as a hypotheses that this influence manifests itself not only in decrease
of level density as compared with excitation of pure fermion system but also in
change of ratio of the mean reduced probabilities of $\gamma$-transitions to the
levels above and below the point of phase transition. It should be noted that due
to the lack of experimental data the authors of the generalized model of 
superfluid nucleus [5] introduced in their model fixed value of the energy of the 
phase transition. This energy corresponds to known value for the infinite and
homogeneous boson system. Experimental data [1] provide the grounds  to consider
this energy for a nucleus as an infinite and inhomogeneous mixture of fermi- 
and bose-systems only as a parameter. Its possible magnitude should approximately two 
times less than that adopted in [5].

All previous experiments on investigation of general picture of the cascade 
$\gamma$-decay below $B_n$ could not reveal this circumstance owing to insufficient 
resolution of the used spectrometers, moreover, in the case when decrease in level
density is compensated by increase in the intensities of transitions populating them.
Calculations of total population $P$ for different $\rho$ and $k$ testify to this 
possibility: relative variation of the population for the tested functional 
dependences changes weaker than for $P-i_1$.
	
  Precise reproduction of the experimental dependence $P-i_1=f(E_{ex})$ cannot be achieved
using a set of standard models for $\rho$ and $k$ [3,4] or values of these parameters
obtained according to method [1]. In the first case one can consider obtained
result as additional argument conforming conclusion [1] about inapplicability
of the models like [4] for predictions of $\rho$ below $B_n$ with the precision
achieved in the experiment. In the second case one should take into account probability
of the dependence $k=\phi(E_\gamma,E_i)$ on not only $\gamma$-quantum energy but also 
excitation energy of the cascade intermediate level $E_i$. Analogous conclusion was 
obtained from both comparison [17] of the experimental and calculated intensities 
of two-step cascades to the levels with excitation energy up to 2 MeV and
comparison [19] of the experimental and calculated total $\gamma$-ray spectra in large
group of nuclei. Qualitatively, from the point of view of theoretical notions of a
nucleus, it is not a surprise: different structure of levels connected by
$\gamma$-transition causes difference in their matrix elements. But the degree of 
influence of this difference on the mean value of matrix element below $B_n$ in a nucleus
can be revealed (and included in expression (1)) only experimentally. At present,
most probably, this can be done only in indirect way by selection of parameters
$\rho$ and $k$ which provide precise reproduction of any known experimental spectra.
 Direct proof of difference  in energy dependences of the $\gamma$-transition strength
functions on the structure of connected levels would require determination
 of their absolute intensities.

Modern experiment does not provide this opportunity. The probability of
considerable strengthening of matrix elements of $\gamma$-transitions in some interval
of excitation energy of the cascade intermediate levels was tested in this work. The
basis for this hypotheses is impossibility to reproduce the data shown in 
figs.~1 and 2 in calculation using the same form of dependence $k=\phi(E_\gamma)$ for the
primary and all following quanta of cascades. 

The lack of experimental information on $\gamma$-transitions depopulating levels
with $2  \leq E_{ex} <B_n$ does not allow one to such practical questions as:

   (a)~how does the local enhancement of matrix element of $\gamma$-transition change
the value of $k$ obtained in analysis [1] for a given energy of decaying level?

   (b)~whether this enhancement lead to redistribution of reduced intensities of
 $\gamma$-transitions to different final levels (in particular, to decrease in 
 $k$ values for final level with energy $0.5 MeV <E_f<2.5$ MeV)?

So, hypotheses of local enhancement of radiative strength functions of secondary
transitions to the levels in the region of ``step-like" structure in level
density allows one not only precisely to calculate cascade population of levels below
3-4 MeV but also to reproduce the dependence $I_{\gamma\gamma}=f(E_1)$ using practically the
same for different nuclei dependences $\rho=\psi(E_{ex})$ 
and  $k=\phi(E_{\gamma})$.

Results presented in this work should be considered as a qualitative description
of the processes occurring in a nucleus. This calculation cannot pretend for
quantitative reproduction of the experimental data cannot also due to the following reasons:

   (a)~impossibility to decrease error in determination the number $N_{2\gamma}$ of the 
observed cascade intermediate levels up to several tens of percent in the experiment
 carried out for single compound state (there is no possibility to exclude or 
estimate correlation of the reduced neutron width with partial radative widths of primary transitions);

   (b)~impossibility to estimate the total cascade population of levels for which
 $i_{\gamma\gamma}$ is lying below the detection threshold or for that intermediate levels for
 which $i_1$ and $i_2$ are unknown;

   (c)~possible inadequacy of the used hypothesis and model notions to the experiment.

Nevertheless, even with these limitations one can conclude that observed cascade
 population can be reproduced only in calculation assuming considerable enhancement
of the radiative strength functions of $\gamma$-transitions to the levels from 
the interval $\simeq 1$ MeV in vicinity of the excitation energy of 3-4 MeV.
\section{
Possibility of experimental test of enhancement of raditive strength functions in the region 0.5$B_n$}\noindent

Direct proof of local enhancement of radiative strength functions for the
transitions to the levels with $E_{ex}\simeq 3$ MeV of even-odd heavy deformed nucleus
requires one to determine reduced relative probability of $\gamma$-transitions from
higher-lying levels in the energy $E_{\gamma}$ interval from several hundreds keV to several 
MeV. This problem cannot be solved by means of classic nuclear spectroscopy using
all types of detectors and methods for determination of the energies of excited levels and their decay modes.

The only realistic way of its solution is experimental measuring of intensity
distributions of the two-step cascades to all possible their final levels up to
the excitation energy 3 MeV and higher of heavy deformed even-odd nucleus.
For even-even lighter nuclei this energy must exceed 4-5 MeV. At the useful statistics 
of several thousands of events for each spectrum $E_1+E_2=B_n-E_f = const$, some part 
of secondary $\gamma$-transitions will be resolved as the pairs of individual peaks.
Analyzing these data by means of the developed methods one can determine intensities
of the secondary $\gamma$-transitions to the levels with $E_f \leq 3-4$ MeV and 
to get a large set of reduced probabilities of $\gamma$-transitions in the region of interest. 

The only reason why were not obtained up to now [14] is that corresponding peaks
with decreasing areas (i.e., peaks with decreasing energy) are located on
increasing Compton background in the sum coincidence spectrum. Potential possibility 
to solve this problem is selection of the cases of simultaneous absorption of
full energy of three successive quanta of cascades to the ground and first excited
state in form of peaks in the sum amplitude spectrum of three coinciding pulses.
Because the energies of the secondary transitions of two-step cascades are known
(and can be determined in the same experiment) then it is possible to get 
distributions of two-step cascades to final levels with $E_f>1$ MeV. This can be done 
also using the systems with several HPGe detectors or pair of such detectors
with suppression of the Compton background. In both cases efficiency must be not 
less than that of detectors in modern ``crystal-balls".
\section{Conclusion}\noindent

The second order phase transition is characterized by abrupt change in properties
of a system as changing its energy. If abrupt enough change in level density 
(i.e., in fact - of nuclear specific heat) was established earlier experimentally
[1] with high enough probability then the results of the performed analysis 
also testify to abrupt change in reduced probability of $\gamma$-transitions
in rather narrow region of nuclear excitations.

The obtained results are independent complementary  confirmation of the existence
in a nucleus of the excitation energy region whre abrupt change of its structure
occurs. Supposedly, there is the transition from the domination of vibrational
excitations to domination of many-quasi-particle excitations. I.e., there is an
analog of the phase transition from the superfluid to usual state of so very
specific system as nucleus.

The data obtained for the excitation energy region of interest should be considered
only as indication to possible existence of such transition. Quantitative
information can be useful for planning of more detailed experiments on investigation
of the problem of interest - dynamics of breaking of Cooper pairs in different
final heterogeneous systems.\\

\newpage
\begin{flushleft}
{\large\bf References}
\end{flushleft}
\begin{flushleft}
\begin{tabular}{@{}r@{ }p{5.65in}}
1.& E.V. Vasilieva, A.M.Sukhovoj, V.A. Khitrov,
Phys. At. Nucl., {\bf 64(2)} (2001) 153\\
& V.A. Khitrov, A.M. Sukhovoj, INDC(CCP)-435, Vienna 21 (2002)\\
& http://arXiv.org/abs/nucl-ex/0110017\\
2. &S.T. Boneva et. al., Nucl.  Phys., (1995) {\bf A589} 293\\
3. &S.G. Kadmenskij, V.P. Markushev, V.I. Furman, Sov. J. Nucl. Phys. {\bf 37}
(1983) 165\\
4.  &W. Dilg, W. Schantl, H. Vonach and M. Uhl, Nucl. Phys., (1973) {\bf A217}
 269\\
5.& E. M. Rastopchin, M. I. Svirin, G. N. Smirenkin, 
Yad. Fiz. {\bf 52}, (1990) 1258\\
 &  E. M. Rastopchin, M. I. Svirin, G. N. Smirenkin, 
 Sov. J. Nucl. Phys., {\bf 52}, (1990) 799\\
6.& A.V. Ignatyuk,  Yu.V. Sokolov
Yad. Fiz.,  {\bf 19} (1974) 1229\\
7.& V.A. Plujko,  Nucl. Phys., {\bf A649} (1999)
209\\\
& V.A. Plujko, S.N. Ezhov, M.O. Kavatsyuk, et al.,\\
 & Journ. Nucl. Sci. Technol.,
Suppl. {\bf 2} (2002) 811\\
& http://www-nds.iaea.or.at/ripl2/\\
8.  &A.M. Sukhovoj, V.A. Khitrov, Instrum. Exp. Tech., {\bf 27} (1984) 1071\\
9. &V.A. Bondarenko et all, Nucl. Phys. {\bf A709} (2002) 1\\
10. &P.Prokofjevs  et all, Nucl. Phys. {\bf A614} (1997) 183\\
11. &V.A. Bondarenko et all, Nucl. Phys. {\bf A619} (1997) 1\\
12.&  Yu.E. Loginov,
 S.Eh. Malyutenkova, Izv RAN, ser. Fiz., to be published\\
13. &http://www-nds.iaea.org/pgaa/egaf.html\\ 
    &G.L. Molnar et al., App. Rad. Isot. (2000) {\bf 53} 527\\
14. &Bondarenko V. A., Honzatko J., Khitrov V. A., Sukhovoj  A. M.,
Tomandl I., \\
&Fizika B (Zagreb) {\bf 11} (2002) 201\\
15. &S.F.~Mughabhab, Neutron Cross Section, vol. 1, part B, Academic Press  1981\\
16. & L.A.Malov and V.G.Soloviev, Yad. Fiz., {\bf 26(4)} (1977) 729\\
17. &A.M.Sukhovoj, V.A. Khitrov,
Phys. At. Nucl., {\bf 64(2)} (2004) 662\\
18. &V. A. Khitrov, Li Chol, A. M. Sukhovoj,
In:  XI International Seminar on \\
&Interaction
 of Neutrons with Nuclei,  Dubna, 28-31 May 2003,\\
&E3-2003-9, Dubna, 2003,
p. 98\\
19.& A.M. Sukhovoj, V.A. Khitrov, E.P. Grigor'ev,
 INDC(CCP), Vienna (2002) {\bf 432} 115
\end{tabular}
\end{flushleft}
\end{document}